\documentstyle[aps,preprint,pra,epsfig]{revtex}

\begin{document}

\draft

\title{Nonlinear Tight-Binding Approximation\\ 
for Bose-Einstein Condensates in a Lattice}

\author{A. Smerzi$^{1,2}$ and A. Trombettoni$^{3}$}
\address{$^1$ 
Istituto Nazionale di Fisica per la Materia BEC-CRS and
Dipartimento di Fisica, Universita' di Trento, I-38050 Povo, Italy\\
$^2$ Theoretical Division, Los Alamos
National Laboratory, Los Alamos, NM 87545, USA\\
$^3$ Istituto Nazionale per la Fisica della Materia and
Dipartimento di Fisica, Universita' di Parma,
parco Area delle Scienze 7A, I-43100 Parma, Italy}

\date{\today}
\maketitle

\begin{abstract}
The dynamics of Bose-Einstein condensates
trapped in a deep optical lattice 
is governed by a discrete nonlinear equation (DNL).
Its degree of nonlinearity
and the intersite hopping rates 
are retrieved from a nonlinear tight-binding approximation
taking into account the effective dimensionality 
of each condensate. We derive analytically the Bloch and the Bogoliubov
excitation spectra, and the velocity of sound waves emitted by 
a traveling condensate.
Within a Lagrangian formalism, we obtain Newtonian-like
equations of motion of localized wavepackets. 
We calculate the ground-state atomic distribution in the presence of an 
harmonic confining potential, and the frequencies of small amplitude  
dipole and quadrupole oscillations. 
We finally quantize the DNL, recovering an extended Bose-Hubbard model. 

\end{abstract}
%\pacs{PACS: 63.20.Pw, 05.45.-a}

%\begin{multicols}{2}

\section{INTRODUCTION} 
Bose-Einstein condensates (BECs) trapped in 
periodic potentials are a new bedtest 
to investigate different and 
fundamental issues of quantum mechanics, 
ranging from quantum phase transitions \cite{greiner02} 
and atom optics 
\cite{rolston02,orzel01} to the dynamics of Bloch  
and Josephson oscillations 
\cite{anderson98,morsch01,cataliotti01}.

The key feature of these systems is the competition/interplay 
between the {\em discrete} translational invariance
(introduced by the periodic potential) 
and the {\em nonlinearity} (due to the interatomic interaction). 
For instance, it has been pointed out that the excitation 
spectrum exhibits a band structure which has analogies 
with the electron Bloch bands in metals
\cite{berg98,javanainen99,chiofalo00}. 
However, the coexistence of  
Bloch bands and nonlinearity 
allows for solitonic structures 
\cite{trombettoni01} and dynamical instabilities 
\cite{wu01,smerzi02,machholm03} 
which do not have an analog neither in metals, nor 
in Galilean invariant nonlinear systems.
   
The BEC in a periodic potential 
is described in mean field (or {\it classical}) approximation
by the Gross-Pitaevskii equation (GPE) Eq.(\ref{GPE}). 
When the interwell barriers of the periodic structure
are high enough (see below), the relevant observables 
of the system are the number of 
particles $N_j(t)$ and the relative phases 
$\varphi_{j+1}(t)-\varphi_{j}(t)$ of the condensates in the 
lattice (the subscript $j$ denotes the different wells of the array). 
In \cite{trombettoni01} it has been shown
that the amplitudes 
$\psi_j=\sqrt{N_j}e^{i \varphi_j}$ satisfy 
a discrete nonlinear Schr\"odinger equation (DNLS), recovered from the 
GPE (\ref{GPE}) in the tight-binding approximation.
Such a mapping has allowed the investigation  
of localized and extended excitations
\cite{trombettoni01,abdullaev01,kalosakas} in the 
framework of the nonlinear lattice theory \cite{hennig99}, and has clarified
the connection between such systems and an array of superconducting
Josephson junctions.
The relative phase dynamics of
different condensates 
has been experimentally investigated 
looking at the interference patterns created by the 
atoms tunneling in the continuum from a vertical lattice \cite{anderson98}, 
or by the overlapping condensates once  
the confining traps are removed \cite{orzel01,cataliotti01,pedri01}. 

In the ``standard" tight-binding approximation, 
the condensate wavefunction localized in the $j$th well is factorized as 
a dynamical part $\psi_j(t)=\sqrt{N_j(t)} e^{i \varphi_j(t)}$ 
and a spatial, real wavefunction $\Phi_j(\vec{r})$
centered in the minimum $\vec{r}_j$ of the well. 
A crucial assumption,  
which will be discussed in detail, is that 
the shape of $\Phi_j(\vec{r})$ does not depend 
on the number of particles $N_j(t)$ in the same well.
Under such conditions, the condensate order parameter $\Psi(\vec{r},t)$ 
can be written as: 
\begin{equation}
\Psi (\vec{r},t)= \sum_{j} \psi_j(t) ~ \Phi_j(\vec{r}).
\label{TB}
\end{equation} 
The DNLS equation is recovered replacing Eq.(\ref{TB}) 
in the GPE Eq.(\ref{GPE}), 
and integrating out the spatial 
degrees of freedom. Neglecting higher order terms, we get \cite{trombettoni01}:
\begin{equation}
\label{DNLS}
i  \hbar \frac{\partial \psi_j}{\partial t} = - K 
(\psi_{j-1} + \psi_{j+1}) + U_2 \mid \psi_j \mid ^2 \psi_j + 
\epsilon_j \psi_j 
\end{equation}
with $K, U_2, \epsilon_j$ 
depending on the geometry of the trapping potentials
and on the average number of atoms in each well [cfr. with 
Eqs.(\ref{u}-\ref{epsilon})]. 

\section{A GENERALIZED TIGHT-BINDING APPROXIMATION} 
In this paper we stress the importance (and exploit the
consequences) to generalize the 
``standard" tight-binding approximation Eq.(\ref{TB}).
This generalization is imposed by the nonlinearity of the GPE 
Eq.(\ref{GPE}), and largely extend the range of
validity of the DNLS Eq.(\ref{DNLS}) in the study of 
the dynamics of weakly coupled BECs. 
The central argument is that the density profile of each condensate
can strongly depend on the number of atoms present at a given instant 
in the same well.
This introduce site- and time- dependent 
parameters in the DNLS Eq.(\ref{DNLS}), modifying, in particular, 
its effective degree of nonlinearity.
Therefore, 
the tight-binding approximation of nonlinear systems has to be generalized as:
\begin{equation}
\label{TB_gen}
\Psi(\vec{r},t)= \sum_{j} \psi_j(t) ~ \Phi_j(\vec{r}; N_j(t)). 
\end{equation}
with $\Phi_j(\vec{r}; N_j(t))$ depending {\it implicitly} on time
through $N_j(t) \equiv |\psi_j(t)|^2$.
We stress here, and discuss again later, that the spatial wavefunctions
$\Phi_j$ (which are considered sufficiently localized in each well) 
can also depend {\it explicitly} on time
due to the excitation of internal modes. 
In the present approach, however, we consider the adiabatic limit in which 
the interwell number/phase dynamics is much slower that the typical 
time associated
with the excitations of such internal modes (and, of course, the cases
where such modes 
are not already present in the initial configuration of the system). 
In this limit, which can is well 
satisfied in typical experiments, the spatial 
wavefunctions in Eq.(\ref{TB_gen}) will adiabatically
follow the tunneling dynamics
and can be approximate with the {\it real} wavefunction 
$\Phi_j(\vec{r}; N_j(t))$. 

\section{THE DISCRETE NONLINEAR EQUATION} 
The BEC dynamics at $T=0$ satisfies the
Gross-Pitaevskii equation (GPE) \cite{dalfovo99}:
\begin{equation}
\label{GPE}
i \hbar \frac{\partial \Psi}{\partial t}= - \frac{\hbar^2}{2 m} 
\nabla^2 \Psi + [V_{ext} + g_0 \mid \Psi \mid^2] \Psi 
\end{equation}
where $V_{ext}$ is the external potential and 
$g_0=\frac{4 \pi \hbar^2 a }{m}$,
with $m$ the atomic mass and $a$ the $s$-wave scattering length: 
$a>0$ ($a<0$) corresponds to an effective interatomic repulsion 
(attraction).
For the sake of clarity, we will focus on the case $a>0$ (as for
$^{87}Rb$ atoms), and with $V_{ext}$ will be given by the 
optical periodic potential $V_{P}$ superimposed to 
a harmonic magnetic field $V_{M}$.
The periodic potential is
\begin{equation}
\label{laser_potential}
V_{P}=V_0 \sin^2{(kx)}  
\end{equation}
where $k=2 \pi / \lambda$ and $\lambda$ is the wavelength of the lasers 
(the lattice spacing is $\lambda / 2$). 
The energy barrier between adjacent sites, $V_0 = s E_R$, 
is expressed in units of the recoil energy $E_R=\frac{\hbar^2 k^2}{2m}$. 
From (\ref{laser_potential}) we see 
that the minima of the laser potential 
are located at the points $x_j=j~\frac{\lambda}{2}$. 
Around these points, $V_{P} \approx 
\frac{m}{2} \tilde{\omega}_x^2 (x-x_j)^2$, where
\begin{equation}
\tilde{\omega}_x=\sqrt{\frac{2 V_0 k^2}{m}}.
\label{omega_tilde}
\end{equation}
The magnetic potential is
$V_M=\frac{m}{2} [\omega_x^2 x^2 + \omega_y^2 y^2 + 
\omega_z^2 z^2]$, with 
$\tilde{\omega}_x>>\omega_x$.
It is convenient to write the external potential as $V_{ext} = V_L + V_D$,
with the confining lattice potential 
$V_L = V_0 \sin^2{(kx)} + \frac{m}{2} [\omega_y^2 y^2 + \omega_z^2 z^2]$,
and the ``driving" field $V_D = \frac{m}{2}~ \omega_x^2~ x^2$. 
$V_D$ has a simple physical meaning:
$F = - \frac{\partial V_D}{\partial x}$ 
is the effective force acting on the center of mass of a condensate wave packet
moving in the periodic potential, see Section VII.

Here we consider a one-dimensional 
optical lattice superimposed to an harmonic driving field,  
but the following considerations can be easily generalized to 
arbitrary $V_D$ and, 
in particular, extended
to the case of two- \cite{greiner01} 
and three- \cite{greiner02} dimensional arrays created by several
counterpropagating laser beams. 

Replacing the nonlinear tight-binding approximation (\ref{TB_gen}) 
in the GPE (\ref{GPE}) and
integrating out the spatial degrees of freedom we find the following DNL: 
\begin{eqnarray}
\label{DNLS_gen}
&& i \hbar \frac{\partial \psi_j}{\partial t} =
\epsilon_j ~ \psi_j  
- \chi ~ [\psi_j(\psi^\ast_{j+1}+\psi^\ast_{j-1})+c.c.]~ \psi_j 
+ \mu_j^{loc}~ \psi_j \cr 
&& - ~ [K+\chi~ (\mid \psi_j \mid^2 + \mid \psi_{j+1} \mid^2)]~ \psi_{j+1} 
   - ~ [K+\chi~ (\mid \psi_j \mid^2 + \mid \psi_{j-1} \mid^2)]~ \psi_{j-1}  
\end{eqnarray}
(we use the normalization $\int d\vec{r}~ \Phi_j^2=1$, 
while the total number of atoms is 
$N_T=\sum_j |\psi_j|^2 \equiv \sum_j N_j$). 
In Eq.(\ref{DNLS_gen}), 
the ``local" chemical potential is the sum of three contributions
\begin{equation}
\mu_j^{loc} = \mu_j^{kin} + \mu_j^{pot} + \mu_j^{int}=
\int d\vec{r} ~ \bigg[ \frac{\hbar^2}{2m}
~(\vec{\nabla} \Phi_j)^2 
+ V_L ~\Phi_{j}^2 
+ g_0 |\psi_j|^2 ~\Phi_j^4 \bigg].  
\label{u}
\end{equation}
$\mu^{loc}_j$ depends on the atom number $N_j$ 
through the condensed wavefunction $\Phi_j(\vec{r}; N_j(t))$. 
The tunneling rates $K_{j,j \pm 1}$ 
between the adjacent sites $j$ and $j \pm 1$ 
also depend, in principle, on the respective populations: 
$K_{j,j \pm 1}(N_j;N_{j \pm 1}) = - \int d\vec{r} ~ \big(\frac{\hbar^2}{2m}
\vec{\nabla} \Phi_j \cdot \vec{\nabla} \Phi_{j \pm 1}
+ \Phi_j V_{ext} \Phi_{j \pm 1})$. In this case, however, we can 
expand the wavefunctions around an average number of atoms per site, $N_0$, 
and keep only the zero 
order term $\Phi_j(N_j) \simeq \tilde{\Phi}_j(N_0)$:
\begin{equation}
K \simeq - \int d\vec{r} ~ \big[ \frac{\hbar^2}{2m}
\vec{\nabla} \tilde{\Phi}_j \cdot \vec{\nabla} \tilde{\Phi}_{j \pm 1}
+ \tilde{\Phi}_j V_{ext} \tilde{\Phi}_{j \pm 1}].
\label{kappa}
\end{equation}
We have checked numerically that higher order terms are 
negligible: for instance, 
with the experimental setup of \cite{cataliotti01},
$V_0=10 E_R$ and $N_0 = 1000$, we have $K_1 = \frac{\partial K}
{\partial N_0} \delta N  \sim 10^{-4} K$.
Similarly, the coefficient $\chi$ is given by:   
\begin{equation}
\chi = - g_0 \int d\vec{r} ~ {\tilde{\Phi}_{j}}^3 \tilde{\Phi}_{j \pm 1}.
\label{epsilon0}
\end{equation}
The on-site energies arising from any 
external potential superimposed to the optical lattice are
\begin{equation}
\epsilon_j = \int d\vec{r}~  V_D~ {\Phi}_j^2;  
\label{epsilon}
\end{equation}
$\epsilon_j \propto j^2$ ($\epsilon_j \propto j$) when the driving field is
harmonic (linear). We notice that in the limit $\tilde{\omega}_x >> \omega_x$
considered here, $\epsilon_j$ does not depend on the on-site atomic 
populations.

In the derivation of Eq.(\ref{DNLS_gen}) we have exploited
the (quasi-)orthogonality of the condensate wavefunctions 
$\int d\vec{r}~ \Phi_j~ \Phi_{j \pm 1} \simeq 0$. Moreover, we have  
verified numerically that spatial integrals 
involving condensates distant more than one site, as well as
terms proportional to
$\int d\vec{r}~\Phi_j^2~\Phi_{j \pm 1}^2$,
can be neglected. 
E.g., with $V_0=10 E_R$ and $N_0 = 1000$,   
$g_0 N_0 \int~ d\vec{r}~ \Phi_j^2~ \Phi_{j \pm 1}^2 / K \sim 10^{-4}$, 
while $\chi N_0/K \sim 10^{-2}$. For $V_0=20 E_R$ and $N_0 = 10000$, 
$\chi N_0/K \sim 10^{-1}$. 
In a double well with e.g. $\tilde{\omega}_x=2 \pi \cdot 100 Hz$ 
and $N_0=10000$, $\chi N_0 \sim K$, while $K_1$ and
$g_0~ N_0~ \int~ d\vec{r}~ \Phi_j^2~ \Phi_{j \pm 1}^2$ can still be ignored. 
For these reasons, we do not neglect 
the $\chi$ terms in Eq.(\ref{DNLS_gen}).
A detailed account of the related numerical study will be presented 
elsewhere.

The atom number dependence in Eq.(\ref{u}) introduces
an effective {\it time-dependent, real,} local
chemical potential $\mu_j^{loc}[N_j(t)]$. 
This reflects an important approximation contained in the DNL:
terms proportional to $\partial \Phi_j/\partial t$ have been neglected. 
In other words, we have neglected the phases associated 
with the spatial dynamics of the  
$\Phi_j(\vec{r}; N_j(t))$ in Eq.(\ref{TB_gen}).
This adiabatic approximation is well satisfied when  
the tunneling time ($\sim N_j / \dot{N}_j$)
is much longer than time scales associated with the change in 
shape of the wavefunctions ($\sim \omega_r^{-1}, \tilde{\omega}_x^{-1}$). 
In this limit, well satisfied in realistic experiments, 
the spatial profile of the wavefunctions adapts adiabatically
to the instantaneous number of atoms present in the respective wells 
\cite{smerzi03}.

The dependence of the local chemical potential on the number of atoms
depends on 
the effective dimensionality of the condensates trapped in each well of the 
lattice. This can be determined 
comparing the interaction chemical potential  
$\mu^{int}_j = |\psi_j|^2 g_0 \int d\vec{r}~
\Phi_j^4$ 
and the three frequencies,
$\tilde{\omega}_x$, $\omega_y$, $\omega_z$ obtained expanding 
the lattice potential around the minima 
of each well 
$V_L \simeq \frac{m}{2} [\tilde{\omega}_x^2 (x-x_j)^2 + \omega_y^2 y^2 +
\omega_z^2 z^2]$.
A sufficiently accurate calculation of
$\mu^{int}_j$ as a function of $N_j$
can be obtained
approximating the condensate order parameters with gaussians 
or Thomas-Fermi functions \cite{nota}. 
Here we first consider some limit cases which are particularly instructive.

When $\tilde{\omega}_x, \omega_y, \omega_z >> \mu^{int}_j$,
the spatial widths of each trapped condensate 
do not depend (in first approximation) 
on the number of particles $N_j$ in the same well, and the 
condensates wavefunctions are well approximated by gaussians.
We consider this as a $0D$ (zero-dimensional) 
case ($nD$, with $n=0,1,2,3$, should not be confused with the 
$spatial$ dimensionality of the lattice),
and the ansatz Eq.(\ref{TB_gen}) reduces to the ordinary 
TBA Eq.(\ref{TB}).
The $1D$ case arises when two frequencies are greater than the interaction
chemical potential. For instance, if 
$\tilde{\omega}_x, \omega_z >> \mu^{int}_j >> \omega_y$, 
the system realizes an array of weakly coupled cigar-shaped 
condensates oriented 
along the $y$-axis: the wavefunction $\Phi_j$ will be factorized as 
a product of two gaussians (in the $x$ and $z$ directions) and a Thomas-Fermi 
in the $y$ variable. In the $2D$ case only one frequency is
smaller than the local interaction chemical potential. If 
$\tilde{\omega}_x >>  \mu^{int}_j >> \omega_y, \omega_z$, we have
an array of pancake-like condensates, with $\Phi_j$ factorized as 
a gaussian (along $x$) and a Thomas-Fermi in the $y$ and $z$ variables 
[see Eqs.(\ref{fact}-\ref{TF})]. The $3D$ case is given by the condition 
$\mu^{int}_j >> \tilde{\omega}_x, \omega_y, \omega_z$ and the wavefunction 
in the $j$th well $\Phi_j$ is simply given by a 
three-dimensional Thomas-Fermi function. 
To summarize:
\begin{displaymath}
\matrix{ 
3D  & \mu^{int}_j~ (\sim N_j^{2/5}) 
>> \omega_a, \omega_b, \omega_c & [spherical]\cr
2D  & \omega_a  >>  \mu^{int}_j~ (\sim N_j^{1/2}) >> \omega_b, 
                                                 \omega_c & [pancake] \cr
1D  & \omega_a,\omega_b >>  \mu^{int}_j~ (\sim N_j^{2/3}) >> \omega_c &
                                                             [cigar] \cr
0D  & \omega_a, \omega_b, \omega_c >>  \mu^{int}_j~ (\sim N_j) 
& [spherical] \cr}
\end{displaymath}
with $a,b,c$ arbitrarly corresponding to the $x,y,z$ spatial directions,
and among square brackets are specified the geometric shapes of the condensates
in each well. 

The crucial point is that the effective
dimensionality of the condensates gives a different scaling of the local 
interaction chemical potential Eq.(\ref{u}) with the number of atoms: 
\begin{equation}
\mu^{loc}_j =  U_{\alpha} \mid \psi_j \mid^{\alpha}
\label{mu_dimensional}
\end{equation}
\begin{equation}
\alpha = {4 \over{2 + D}}, ~~D=0,1,2,3.
\label{alpha}
\end{equation}
where 
$U_{\alpha}$ is a constant which does not depend on the number of atoms nor 
on the site index.
In the following we will often consider, for the sake
of clarity, the limit cases 
when the local chemical potential is given by Eq.(\ref{mu_dimensional})
(generalization to more complicate functional dependences of $\mu_j^{loc}$
from $N_j$ are straightforward).
The DNLS Eq.(\ref{DNLS}) is recovered from the DNL Eq.(\ref{DNLS_gen})
in the case $D=0$ (i.e. $\alpha=2$) and neglecting terms proportional 
to $\chi$. 

The derivation of the Hamiltonian of the system requires some care.
The dynamical variables $\psi_j^\ast, i \hbar \psi_j$ are canonically conjugate
($\dot{\psi_j} =\frac{\partial {\cal{H}}_{eff}} {\partial 
(i \hbar \psi_j^{\ast})}$)
with respect to the effective Hamiltonian  
\begin{eqnarray}
\label{HAM_eff}
{\cal{H}}_{eff} =&& {\sum_j} \Bigg\{\epsilon_j  \psi^\ast_j \psi_{j} - 
K (\psi_j^\ast \psi_{j+1}+c.c.) \cr
&&  - \chi \big[ \mid \psi_j \mid^2 \psi_j 
(\psi^{\ast}_{j+1}+\psi^{\ast}_{j-1}) + c.c.\big] + 
{2 \over {2 + \alpha}} U_{\alpha} |\psi_j|^{\alpha + 2} \Bigg\}
\end{eqnarray}
(with the nonlinear term 
${2 \over {2 + \alpha}} U_{\alpha} |\psi_j|^{\alpha + 2} $ obtained from 
$\psi_j \int d \psi_j^\ast \mu_j^{loc}$).

The effective Hamiltonian ${\cal{H}}_{eff}$ 
is an exact integral of motion, but differs from the
``adiabatic" Hamiltonian retrieved simply replacing Eq.(\ref{TB_gen})
in the Gross-Pitaevskii energy functional:
\begin{eqnarray}
\label{HAM_ad}
{\cal{H}}_{ad} =&& {\sum_j} \Bigg\{\epsilon_j \psi^\ast_j \psi_{j} - 
K (\psi_j^\ast \psi_{j+1}+c.c.) + \cr
&&  - \chi \big[ \mid \psi_j \mid^2 \psi_j 
(\psi^{\ast}_{j+1}+\psi^{\ast}_{j-1}) + c.c.\big] + 
{1 \over 2} U |\psi_j|^4 \Bigg\},
\end{eqnarray}
with $U = g_0 \int d\vec{r} ~\tilde{\Phi}_j^4$.
The ${\cal{H}}_{eff}$ and ${\cal{H}}_{ad}$ are identical only in the
$0D$ case. In general, ${\cal{H}}_{ad}$
is not exactly, but only ``adiabatically", conserved during the dynamics.

\section{EXCITATION SPECTRA} 
We now derive the Bloch excitation spectra and the Bogoliubov 
dispersion relation of the system 
(with $\epsilon_j = 0$), calculate the sound velocity
and investigate the dynamical stability of condensate traveling waves. 
Eigenfunctions of the DNL are the plane waves 
$\psi_n= \psi_0~e^{i(p n- \mu t/\hbar)}$,
with chemical potential and energy per site:
\begin{eqnarray}
\label{chempot}
\mu &=& \mu^{loc} - 2~(K + 4~ \chi~N_0) \cos p \cr
E &=& E^{loc} - 2~(K + 2~ \chi~N_0) N_0 \cos p
\end{eqnarray}
where $N_0 = \mid \psi_0 \mid^2$, 
$\mu^{loc} = \mu^{loc}_j|_{\psi_j=\psi_0}=U_\alpha \mid \psi_0 \mid^\alpha$, 
and 
$E^{loc} = \psi_0 \int d \psi_j^\ast \mu_j^{loc}|_{\psi_j=\psi_0}=
2 U_\alpha \mid \psi_0 \mid^{\alpha+2}/(\alpha+2)$ 
[see Eq.(\ref{mu_dimensional})].
From Eq.(\ref{chempot}) we can recover the group velocity 
of Bloch waves with quasimomentum $p$: 
$v_g \equiv {\frac{1}{N_0}} {\frac{\partial E}{\partial p}} = 
2~(K + 2~ \chi~N_0) \sin p$.

We remark that the Bloch energy $E$ and the chemical potential 
$\mu$ have the same 
$\cos p$ dependence on the quasimomentum $p$,
but with different coefficients. 
This introduces different effective masses 
for the system (see also \cite{menotti1}),
which will enter in peculiar ways in the equations discussed in this paper, 
as will be reported elsewhere \cite{menotti}. 
Here we will write down our results only in terms of the 
DNL parameters.  

In order to derive the Bogoliubov dispersion relation of the system, 
we perturb the large amplitude wave as
$\psi_n=[\psi_0+u(t) e^{iqn}+v^{\ast}(t) e^{-iqn}]~
e^{i(p n-\mu t / \hbar)}$. Retaining only first order terms proportional 
to $u/\psi_0$ and $v/\psi_0$, we get
\begin{equation}
i \hbar \frac{d}{dt} 
\left( \matrix{ 
         u  \cr
         v  \cr } \right)=
\left( \matrix{ 
a+b & c \cr
-c^\ast & a-b } \right)
\left( \matrix{ 
         u  \cr
         v  \cr } \right) = \omega_{\pm} \left( \matrix{
         u  \cr
         v  \cr } \right)
\label{gen_alpha}
\end{equation}
with 
$a= 2 (K + 4 \chi N_0) \sin p \sin q$, 
$b= 2 K \cos p - 
2 (K + 4 \chi N_0) \cos p \cos q + 
N_0 \frac{\partial \mu^{loc}} {\partial N_0}$ and 
$c=-4 \chi \psi_0^2 \cos p \, 
(1 + \cos q) +\psi_0^2 \frac{\partial \mu^{loc}} {\partial N_0}$. 
Up to the order $\chi^2 N_0^2 / K^2$, we get the eigenvalues:
\begin{eqnarray}
&& \omega= 2 (K + 4 \chi N_0) \sin{p} \sin{q} \pm   \cr 
&& 2 \sqrt{4 K (K + 8 \chi N_0) \cos^2{p} \sin^4{\frac{q}{2}} +
2 (K + 2 \chi N_0) \frac{\partial \mu}{\partial N_0} N_0 \cos{p} 
\sin^2{q \over 2} }.
\label{mi_alpha}
\end{eqnarray}

\section{SOUND-WAVES AND INSTABILITIES}

The small $q$ (large wavelength) limit of the Bogoliubov dispersion relation
Eq.(\ref{mi_alpha}) is linear. Therefore,
the system support (low amplitude) sound waves (propagating on top of the
large amplitude traveling wave $\psi_0~e^{i(p n- \mu t/\hbar)}$)
having velocity
\begin{equation}
v_s = \frac{\partial \omega}{\partial q}|_{q = 0} =
2 (K + 4 \chi N_0) \sin{p} \pm  \sqrt{
2 (K + 2 \chi N_0) \frac{\partial \mu}{\partial N_0} N_0 \cos{p}  }
\label{sound}
\end{equation}
with $\mu$ given by Eq.(\ref{chempot}).
The $+~(-)$ sign corresponds to a sound wave propagating in the same (opposite)
direction of the large amplitude traveling wave.
Notice that, contrary to the case of a Galilean invariant system 
($V_{0} = 0$),
the sound velocity depends on the quasimomentum $p$. Moreover, $v_s$ depends
on the effective dimensionality of the condensates, since
(from Eq.(\ref{mu_dimensional},\ref{alpha}))
$\frac{\partial \mu}{\partial N_0} N_0 \sim \alpha~ U_\alpha~ N_0^{\alpha/2}$.

In the limit $V_0 = 0$, the system is
energetically unstable if $\omega < 0$, namely when the group velocity 
is larger than the sound velocity (Landau criteria for superfluidity).
This instability is present also when the the system has a discrete 
translational invariance ($V_0 > 0$): from the Bogoliubov excitation spectrum 
Eq.(\ref{mi_alpha}) and the condition $\omega < 0$, we have that the system 
is not superfluid when 
\begin{equation}
[2 (K + 4 \chi N_0) \sin{p}]^2 >   
2 (K + 2 \chi N_0) \frac{\partial \mu}{\partial N_0} N_0 \cos{p}  
\label{ei}
\end{equation}

There is a further, different (dynamical) instability mechanism, 
which disappears in the translational invariant limit 
(when $a > 0$). This instability is associated with 
the appearance of an imaginary component in the Bogoliubov frequencies:
from Eq.(\ref{mi_alpha}), this component appears if $\cos p < 0$.
This reflects on an exponential increase of the amplitude
of the perturbation modes, with the consequent 
strong dephasing and energy dissipation of the
condensate traveling wave.
The unstable modes $q$, for a given quasimomentum $p$, are given by: 
\begin{equation}
2 \, \bigg( 1+\frac{6 \chi N_0}{K} \bigg) \mid \cos p \mid \sin^2{q \over 2} < 
\frac{\partial \mu}{\partial N_0} N_0. 
\label{mi_alpha_im}
\end{equation} 
For $\alpha=2$ and $\chi=0$ we recover the standard DNLS results 
\cite{kivshar92,smerzi02}. The onset of energetic and dynamical instabilities 
with an arbitrary $V_0$ has been studied numerically in \cite{wu01}.
Experimental evidences are 
reported on \cite{cataliotti}. A different manifestation of 
the instability is associated with the self-trapping of a condensate
wavepacket at rest in an optical lattice \cite{trombettoni01}. 
First experimental results are reported in \cite{morsch02}.

\section{GROUND-STATE ATOMIC DISTRIBUTION} 
We now consider a magnetic harmonic potential 
superimposed to the optical lattice
$\epsilon_j=\Omega j^2$, with 
$\Omega=m\omega_x^2 \lambda^2/8$. For a large nonlinearity,
the ground-state atomic distribution can be calculated from the DNL 
(\ref{DNLS_gen}) in Thomas-Fermi approximation, 
i.e. neglecting the kinetic terms proportional to $K$ and $\chi$ 
with respect to the nonlinear term:
\begin{equation}
N_j=\bigg(\frac{\nu - \Omega j^2}
{U_{\alpha}}\bigg)^{2/\alpha}=
\bigg(\frac{\nu}{U_\alpha}\bigg)^{2/\alpha}
\bigg(1-\frac{j^2}{j_{inv}^2}\bigg)^{2/\alpha}
\label{gs}
\end{equation} 
where the inversion point is $j_{inv}^2=\nu/\Omega$.
Replacing sums with integrals, we get 
\begin{equation}
\nu=\bigg( 
\frac{N_T \Omega^{1/2} U_{\alpha}^{2/\alpha}}{C_{\alpha}} 
\bigg)^{2 \alpha/(\alpha+4)}
\label{nu}
\end{equation}
where $C_{\alpha}=2^{4/\alpha+1} 
[\Gamma(2/\alpha+1)]^2 / \Gamma{(4/\alpha+2)}$ 
is a numerical constant ($\Gamma$ is the Gamma function). 
For $\alpha=2$ ($0D$), $C_{2}=4/3$, while
for $\alpha=1$ ($2D$), $C_{1}=16/15$.

\section{NEWTONIAN DYNAMICS AND SMALL AMPLITUDE OSCILLATION FREQUENCIES} 
We now study the wave-packet dynamics of a BEC in an optical lattice.
We resort to a variational approach, 
previously considered in \cite{trombettoni01}. 
Here we use a general variational wavefunction
\begin{equation}
\psi_j=\sqrt{\cal K(\sigma)} f\bigg(\frac{j-\xi}{\sigma}\bigg) 
e^{ip(j-\xi)+i \frac{\delta}{2}(j-\xi)^2}
\label{var}
\end{equation}  
where $\xi(t)$ and $\sigma(t)$ are, respectively, 
the center and the width of the wavepacket, $p(t)$ and $\delta(t)$ their 
associated momenta and $\cal K(\sigma)$ a normalization factor 
(such that $\sum_j N_j=N_T$). $f$ is a generic function, even in the 
variable $X=(j-\xi)/\sigma$. E.g., we can choose $f(X)=e^{-X^2}$ 
or $f(X)=(1-X^2)^{1/\alpha}$ (with $-1 \le X \le 1$) to describe, 
respectively, the dynamics of a gaussian or 
a Thomas-Fermi wave packet. With the Lagrangian 
${\cal{L}} =i \hbar \sum_j \psi_j^{\ast} \dot{\psi}_j - 
{\cal{H}}_{eff}$ 
we can recover the equations of motions for the variational 
parameters $q_i(t) \equiv \xi(t), \sigma(t), p(t), \delta(t)$, given by
$\frac{d}{d t} \frac{\partial {\cal{L}}}{\partial \dot{q}_i}=
\frac{\partial {\cal{L}}}{\partial q_i}$.
With the variational wavefunction Eq.(\ref{var}), the Lagrangian becomes:
\begin{eqnarray}
\label{lagran}
\frac{{\cal L}}{N_T}= \hbar p \dot{\xi} - \hbar \sigma^2 \dot{\delta} 
\frac{{\cal I}_2}{2 {\cal I}_1}
-V_D(\xi, \sigma)
-\tilde{U}_{\alpha} \frac{N_T^{\alpha/2}}{\sigma^{\alpha/2}}+ 
\frac{2 K}{{\cal I}_1}{\cal I}_J(\sigma;\delta) \, \cos{p}+ 
\frac{2 \chi N_T}{\sigma {\cal I}_1^2}{\cal I}_{\chi} (\sigma;\delta) \, 
\cos{p}  
\end{eqnarray}
where $V_D(\xi, \sigma)=\frac{1}{{\cal I}_1} 
\int dX f^2(X) \epsilon(\sigma X + \xi)$, 
$\tilde{U}_{\alpha}=2 {U}_{\alpha} {\cal I}_{NL}/[(\alpha+2) 
{\cal I}_1^{\alpha/2+1} ]$, ${\cal I}_J(\sigma;\delta)=\int dX f(X+1/2\sigma) 
f(X-1/2\sigma) e^{i \sigma \delta X}$ and 
${\cal I}_{\chi} (\sigma;\delta)=\int dX f(X+1/2\sigma) 
f(X-1/2\sigma) [f^2(X+1/2\sigma)+f^2(X-1/2\sigma)] e^{i \sigma \delta X}$. 
Furthermore ${\cal I}_1=\int dX f^2(X)$, ${\cal I}_2=\int dX X^2 f^2(X)$ 
and ${\cal I}_{NL}=\int dX f^{\alpha+2}(X)$ are real numbers which 
depend on the particular choice of $f$. 
From the Lagrangian equation of motion we get the group velocity  $\dot{\xi}$
and the effective force acting on the center of mass of the wavepacket:
\begin{eqnarray}
\label{group}
\hbar \dot{\xi} &=& [\frac{2 K}{{\cal I}_1}{\cal I}_J(\sigma;\delta) +
\frac{2 \chi N_T}{\sigma {\cal I}_1^2}{\cal I}_{\chi} (\sigma;\delta)] \,
\sin{p} \cr
\hbar \dot{p} &=& - \frac {\partial V_D}{\partial \xi}
\end{eqnarray}
The frequency of small amplitude oscillations of the wavepacket 
driven by an harmonic field $\epsilon_j=\Omega j^2$ (which gives 
$V_D(\xi,\sigma) = \Omega (\xi^2+\sigma^2 \frac{ {\cal I}_2 } 
{ {\cal I}_1 })$), is: 
\begin{equation}
\omega_{dip}^2=\frac{2 \Omega}{\hbar^2} 
\bigg( 2 K+ \frac{8 \chi N_0 {\cal I}_3}
{{\cal I}_1^2} \bigg) 
\label{omega-dip}
\end{equation}
where $N_0=N_T/2 \sigma$ \cite{nota-gauss}.
Eq.(\ref{omega-dip}) has been calculated in 
the limit of a large width $\sigma >> 1$, 
where ${\cal I}_J(\sigma;0) \simeq {\cal I}_1$ and 
${\cal I}_\chi(\sigma;0) \simeq 2 {\cal I}_3$, with 
${\cal I}_3= \int dX f^4(X)$.  
The same results follow from the exact equation of motion for 
$\xi=\sum_j j N_j$ and $p=\varphi_{j+1}-\varphi_{j}$, 
with the latter assumed equal for each $j$ along the array 
(and using the fact that $\sum_j \sqrt{N_j N_{j+1}} \simeq N_T$ and 
$\sum_j N_j \sqrt{N_j N_{j+1}} \simeq \sum_j N_j^2)$.
For $\chi=0$ Eq.(\ref{omega-dip}) coincides with the result in 
\cite{cataliotti01}.

To calculate the quadrupole oscillation frequency we need the equation of
motion for the width $\sigma$ and the 
conjugate momentum $\delta$ (still with $V_D = \Omega \xi^2$):
\begin{eqnarray}
\label{quad}
-\hbar \dot{\sigma} \frac{{\cal I}_2}{{\cal I}_1} &=& 
\frac{2K}{\sigma {\cal I}_1}\frac{\partial I_J}{\partial \delta} \cos{p}+
\frac{2 \chi N_T}{\sigma^2 {\cal I}_1^2} 
\frac{\partial I_\chi}{\partial \delta} \cos{p}         \cr
\hbar \dot{\delta} \frac{{\cal I}_2}{{\cal I}_1} &=& -2 \Omega \frac{{\cal I}_2}{{\cal I}_1}
+ \frac{\alpha \tilde{U}_{\alpha} N_T^{\alpha/2}}{2 \sigma^{\alpha/2+2}}+
\frac{2K}{\sigma {\cal I}_1}\frac{\partial I_J}{\partial \sigma} \cos{p} +
\frac{2 \chi N_T}{\sigma^2 {\cal I}_1^2} 
\bigg( \frac{\partial I_\chi}{\partial \sigma} -
\frac{I_\chi}{\sigma} \bigg) \cos{p}
\end{eqnarray}
The equilibrium position is given by $\dot{\delta}=0$, 
$\dot{\sigma}=0$, $\xi=0$ and $p=0$. 
Linearizing around the equilibrium for the Thomas-Fermi ground-state 
(\ref{gs}), 
and after a lenghty calculation, we get the frequency 
of the quadrupole oscillations:
\begin{equation}
\omega_{quadr}^2=\frac{\Omega \alpha (\alpha+4) {\cal I}_{NL}}
{2\hbar^2 {\cal I}_2 (\alpha+2)} 
\bigg( 2 K+ \frac{8 \chi N_0 {\cal I}_4 }
{{\cal I}_1 {\cal I}_2} \bigg)
\label{omega-quadr-TF}
\end{equation} 
where ${\cal I}_4=\int dX X^2 f^4(X)$ \cite{nota-cost}.  
Eq.(\ref{omega-quadr-TF}) shows that the quadrupole 
frequency explicitely depends on the effective dimensionality 
of the condensates in each well \cite{nota-N}. 
Collecting Eqs.(\ref{omega-dip}) 
and (\ref{omega-quadr-TF}) we get
\begin{equation}
\frac{\omega_{quadr}^2}{\omega_{dip}^2}=
\frac{\alpha (\alpha+4) {\cal I}_{NL}}{4(\alpha+2){\cal I}_2} \cdot
\frac{1+4 \frac{\chi N_0}{K} \frac{ {\cal I}_4}{ {\cal I}_1 {\cal I}_2}}
{1+4 \frac{\chi N_0}{K} \frac{ {\cal I}_3}{ {\cal I}_1^2}}.  
\label{rapp-omega}
\end{equation}
When $\chi N_0 << K$,  
$\frac{\omega_{quadr}^2}{\omega_{dip}^2}= 2~ {{D + 3} \over {D + 2}}$. 
In particular, $\omega_{quadr}^2/\omega_{dip}^2=3$ in the zero-dimensional 
case, $D=0$. The $2D$, $\chi =0$ result, 
$\omega_{quadr}^2 / \omega_{dip}^2= 5 / 2$, 
is in agreement with the results of \cite{kramer02}. 

\section{NUMERICAL ESTIMATES}
We now consider a specific example to further 
clarify the calculation of the 
DNL coefficients Eqs.(\ref{u}-\ref{epsilon}).
Considering the experimental apparatus of
\cite{cataliotti01}, we put: $\omega_x=2\pi \times 9 Hz$, 
$\omega_y=\omega_z \equiv \omega_r=2\pi \times 90Hz$,
$\lambda=795$ nm, $E_R/h=3.6 \, kHz$ and
$V_0 / E_r$ from $2$ to $15$.
From Eq.(\ref{omega_tilde}), we obtain
$\tilde{\omega}_x/2\pi=\sqrt{s} \cdot 7.2 \, kHz$.

Since $\tilde{\omega}_x >> \omega_r$, we find 
$\mu^{kin}_j +\mu^{harm}_j \approx 
\hbar \tilde{\omega}_x/2$. With an average value of atoms in each well 
$N_0 \sim 1000$ and with
$V_0=5 E_r$, we obtain an interaction chemical potential
$\mu_j^{int}  \sim h \cdot 2 kHz$, which corresponds,
according to the table in Section III, 
to the $2D$ case. The system can be seen as a horizontal pile of pankakes, 
having a smaller diameter at the border of the pile, 
dense at the center and more dilute at the surface. 
In this limit we have:
\begin{equation}
\Phi_j(\vec{r},N_j(t)) \simeq \phi_G^{(j)}(x-x_j) \phi_{TF}^{(j)}(y,z)   
\label{fact}
\end{equation}
where $\phi_G^{(j)}=(\sigma \sqrt{\pi})^{-1/2} e^{-(x-x_j)^2/2\sigma^2}$ 
is a gaussian with width $\sigma$ (we impose
$\int dx (\phi_G^{(j)})^2=\int dy dz (\phi_{TF}^{(j)})^2=1$).
A variational calculation shows that there is a very weak dependence
of $\sigma$ on $N_j$; we therefore
assume it as site-independent \cite{pedri01,trombettoni01-thesis}:
$\sigma=\frac{\lambda}{2 \pi s^{1/4}}$. 
Replacing Eq.(\ref{fact}) in Eq.(\ref{GPE})
and integrating out along the $x$ direction, we obtain an equation for
$\phi_{TF}^{(j)}(y,z)$:
\begin{equation}
\label{GPE-2D}
[- \frac{\hbar^2}{2 m} 
\nabla_{\vec{R}}^2 + {\cal V}(\vec{R})  
+ \tilde{g_0} N_j (\phi_{TF}^{(j)})^2] \phi_{TF}^{(j)}=
\mu^{int}_j \phi_{TF}^{(j)} 
\end{equation}
with $\tilde{g_0}=g_0/\sqrt{2 \pi} \sigma$; $\vec{R}=(y,z)$ is the vector 
expressing the position in the $y$-$z$ radial plan and 
${\cal V}(\vec{R})=\frac{m}{2}\omega_r^2 R^2$. 
In Thomas-Fermi approximation (i.e. neglecting the kinetic terms in 
Eq.(\ref{GPE-2D})), we find 
\begin{equation}
\phi_{TF}^{(j)}(\vec{R})=\bigg(\frac{\mu^{int}_j-{\cal V}(\vec{R})}
{\tilde{g_0} N_j}\bigg)^{1/2}. 
\label{TF}
\end{equation}
The inversion point is $R_\perp^2=2 \mu^{int}_j/m\omega_r^2$. 
Replacing Eq.(\ref{TF}) in Eqs.(\ref{u}) we obtain 
\begin{equation}
\mu_j^{loc}=\sqrt{\frac{m \omega_r^2 g_0}{\sqrt{2 \pi} 
\pi \sigma}} N_j^{1/2}.
\label{u-2D}
\end{equation} 
The on-site energies (\ref{epsilon}) are given by $\epsilon_j=\Omega j^2$, 
where $\Omega=\frac{m}{2} m \omega_x^2 (\frac{\lambda}{2})^2$. 
We have neglected the kinetic terms 
$\epsilon_j^{(kin)}= \frac{\hbar^2}{2m} \int d\vec{R} 
(\vec{\nabla}_R \phi_{TF}^{(j)})$, consistent with the
the Thomas-Fermi approximation (\ref{TF}).
Using Eqs.(\ref{u-2D}) we get 
the DNL (\ref{DNLS_gen}) with $D=2~(\alpha=1)$ and 
[see Eq.(\ref{mu_dimensional})]
\begin{equation}
U_1=\sqrt{\frac{m \omega_r g_0}{\sqrt{2 \pi} \pi \sigma}}. 
\label{u-grande-2D} 
\end{equation}

The population distribution in the ground state, according to Eq.(\ref{gs}), 
is given by 
\begin{equation}
N_j=
\bigg(\frac{\nu}{U_1}\bigg)^2 \bigg(1-\frac{j^2}{j_{inv}^2}\bigg)^2.
\label{gs_2D}
\end{equation}
The inversion point is $j_{inv}=\sqrt{\frac{\nu}{\Omega}}$ and 
the discrete chemical potential (\ref{nu}) is 
is $\nu=(15 N_T U_1^2 \sqrt{\Omega}/16)^{2/5}$. Therefore
\begin{equation}
j_{inv}^2=\frac{2 \hbar \bar{\omega}}
{m \omega_x^2 d^2} \bigg(\frac{15}{8\sqrt{\pi}} N_T 
\frac{a d}{a_{ho} \sigma} \bigg)^{2/5}
\label{j_inv_2D}
\end{equation}
where 
$d=\lambda/2$, $a_{ho}=\sqrt{\hbar/m \bar{\omega}}$ and 
$\bar{\omega}=(\omega_r^2 \omega_x)^{1/3}$. The $D=2$ ground state
Eqs.(\ref{gs_2D}-\ref{j_inv_2D}) is 
in agreement with \cite{pedri01}, previously calculated with a 
different approach. 

\section{QUANTUM CASE: AN EXTENDED BOSE-HUBBARD MODEL} 
The quantization of the DNL equation requires some care.
The quantum equation for the bosonic gas in an external potential is:
\begin{equation}
i \hbar \frac{\partial}{\partial t}\hat{\Psi} (\vec{r},t) = 
[T+V_{ext}+g_0 \hat{\Psi}^{\dag} \hat{\Psi} ] \hat{\Psi},
\label{quantum}
\end{equation} 
The Gross-Pitaevskii equation (\ref{GPE}) can be retrieved  
introducing the classical field $\Psi=\langle \hat{\Psi} \rangle$ and
with $\langle \hat{\Psi}^{\dag} \hat{\Psi}  \hat{\Psi} \rangle 
\simeq \langle \hat{\Psi}^{\dag} \rangle \langle \hat{\Psi} \rangle
\langle \hat{\Psi} \rangle $.

In tight-binding approximation
\begin{equation}  
\label{QTB}
\hat{\Psi}(\vec{r},t)= \sum_j \hat{\psi}_j(t) 
\Phi_j(\vec{r})
\end{equation}
(with $\hat{\psi}^{\dag}_j \hat{\psi}_j$ the bosonic number operator), 
we obtain the Bose-Hubbard model (BHM) \cite{fisher89,jaksch98} 
\begin{equation}
\hat{H}={\sum_j} \big\{ - K (\hat{\psi}^{\dag}_j \hat{\psi}_{j+1} + h.c.)
+ {U_2 \over 2} 
(\hat{\psi}^{\dag}_j \hat{\psi}^{\dag}_j \hat{\psi}_j\hat{\psi}_j) + 
\epsilon_j \hat{\psi}^{\dag}_j \hat{\psi}_j \big\}
\label{B-H}
\end{equation}
We now discuss the case in which the localized wavefunction $\Phi_j$ 
in the $j$th well adiabatically depends on the average number of particles 
in the same well: the generalization to the quantum case of (\ref{TB_gen}) is 
\begin{equation}
\label{QTB_gen}
\hat{\Psi}(\vec{r},t)= \sum_{j} \hat{\psi}_j(t) 
\Phi_j(\vec{r}; N_j(t))
\end{equation}
where
\begin{equation}
\label{N_ave}
N_j  = \langle \hat{\psi}_j^\dagger \hat{\psi}_j \rangle.
\end{equation}
Replacing the ansatz (\ref{QTB_gen}) in (\ref{quantum}), it is easy to recover
the quantum equation of motion for bosonic operators $\hat{\psi}_j$.
Such equations are generated, with the standard bosonic commutation relations,
from the extended Bose-Hubbard Hamiltonian:
\begin{eqnarray}
\label{HAM_quan}
\hat{H} = && {\sum_j} \Bigg\{ \epsilon_j \hat{\psi}^\dagger_j \hat{\psi}_{j} + 
{1 \over 2} U 
(\hat{\psi}^\dagger_j \hat{\psi}^\dagger_j \hat{\psi}_j \hat{\psi}_{j}) \cr
&& - K (\hat{\psi}^{\dag}_j \hat{\psi}_{j+1} + h.c.) 
- \chi [\hat{\psi}^\dag_{j} \hat{\psi}_{j}  \hat{\psi}_{j} 
(\hat{\psi}^\dag_{j+1} + \hat{\psi}^\dag_{j-1}) \, + \, 
h.c.]  \Bigg\}
\end{eqnarray}
with the parameter $K, \chi, \epsilon_j, U$ expressed as in the
classical DNL Eq.(\ref{DNLS_gen}). Notice that the extended BHM can be 
alternatively recovered quantizing the classical {\it adiabatic} Hamiltonian
${\cal{H}}_{ad}$ Eq.(\ref{HAM_ad}) (and not the {\it effective} Hamiltonian
Eq.(\ref{HAM_eff})). 

\section{CONCLUSIONS} The Gross-Pitaevskii
dynamics of a Bose-Enstein condensate
trapped in a deep periodic potential can be studied in terms of a
discrete, nonlinear equation. This mapping allows a clear and intuitive picture
of the main dynamical properties of the system, which can be 
calculated analytically. We have shown that 
the slopes of the 
energy and chemical potential Bloch excitation spectra,
with respect to the quasimomentum of the condensate, are different.
We have calculated the Bogoliubov dispersion relation, and studied
the sound-wave velocity as a function 
of i) the effective dimensionality of each condensate,
and ii) the quasimomentum of the carrier wave. 
Through a Lagrangian formalism, we have 
recovered Newtonian-like equation of motion of localized wavepackets, 
and the frequencies of 
dipole and quadrupole small amplitude oscillations.
We have finally quantized the discrete nonlinear Hamiltonian recovering 
an extended Boson-Hubbard model. 

{\it Note added in proof:} An equation similar to the DNL (\ref{DNLS_gen}) 
(with $\alpha=2$, and including the term proportional to 
$\int d\vec{r} ~ {\tilde{\Phi}_{j}}^2 \tilde{\Phi}_{j \pm 1}^2$) 
has been derived by \"Oster, Johansson, and Eriksson \cite{oster03} 
to describe the amplitude of an electric field in an array of coupled 
waveguides embedded in a material with Kerr nonlinearities.

{\bf Acknowledgements.}
We thank C. Menotti for several valuable comments. 
We aknowledge dicussions with 
S. Giorgini, M. Kr\"amer, L. P. Pitaevskii, and S. Stringari. 
A.T. thanks the CRS-BEC of Trento, where part of this work was completed, 
for the kind hospitality. 
This work has been partially supported by the DOE.

%\end{multicols}{2}


\begin{thebibliography}{10}

\bibitem{greiner02} M. Greiner, O. Mandel, T. Esslinger, 
T.W. H\"ansch, and I. Bloch, Nature {\bf 415}, 39 (2002).

\bibitem{rolston02} S.L. Rolston and W.D. Phillips, Nature {\bf 416}, 
219 (2002) and ref.s therein.

\bibitem{orzel01} C. Orzel, A.K. Tuchman, M.L. Fenselau, 
M. Yasuda, and M.A. Kasevich, Science {\bf 291}, 2386 (2001) and 
ref.s therein.

\bibitem{anderson98} B.P. Anderson and M.A. Kasevich,
Science {\bf 282}, 1686 (1998).

\bibitem{morsch01} O. Morsch, J.H. M\"uller, M. Cristiani, D. Ciampini and
A. Arimondo, Phys. Rev. Lett. {\bf 87}, 140402 (2001).

\bibitem{cataliotti01} F.S. Cataliotti, S. Burger, C. Fort, 
P. Maddaloni, F. Minardi, A. Trombettoni, A. Smerzi, 
and M. Inguscio, 
Science {\bf 293}, 843 (2001). 

\bibitem{berg98} K. Berg-S{\o}rensen and K. Molmer, 
Phys. Rev. A {\bf 58}, 1480 (1998).

\bibitem{javanainen99} J. Javanainen, Phys. Rev. A {\bf 60}, 4902 (1999).

\bibitem{chiofalo00} M.L. Chiofalo and M.P. Tosi, Phys. Lett. {\bf A268},
406 (2000).  

\bibitem{trombettoni01} A. Trombettoni and A. Smerzi, 
Phys. Rev. Lett. {\bf 86}, 2353 (2001).

\bibitem{wu01} B. Wu and Q. Niu, Phys. Rev. A {\bf 64}, 061603(R) (2001).

\bibitem{smerzi02} A. Smerzi, A. Trombettoni, P.G. Kevrekidis, 
and A.R. Bishop, Phys. Rev. Lett. {\bf 89}, 170402 (2002).

\bibitem{machholm03} M. Machholm, C.J. Pethick, and H. Smith, 
Phys. Rev. A {\bf 67}, 053613 (2003). 

\bibitem{abdullaev01} F.K. Abdullaev, B.B. Bazaikov, 
S.A. Darmanyan, V.V. Konotop, and M. Salerno, Phys. Rev. A {\bf 64}, 
043606 (2001); G.L. Alfimov, P.G. Kevrekidis, V.V. Konotop and M. Salerno,
 Phys. Rev. E {\bf 66}, 046608 (2002).

\bibitem{kalosakas} G. Kalosakas, K.\O. Rasmussen, and A.R. Bishop,
Phys. Rev. Lett. {\bf 89}, 030402 (2002). 

\bibitem{hennig99} See e.g. D. Hennig and G.P. Tsironis, Phys. 
Rep. {\bf 307}, 333 (1999), and ref.s therein. 

\bibitem{pedri01} P. Pedri, L.P. Pitaevskii, S. Stringari,
S. Burger, F.S. Cataliotti, S. Burger, C. Fort,
P. Maddaloni, F. Minardi, and M. Inguscio,
Phys. Rev. Lett. {\bf 87}, 220401 (2001).

\bibitem{dalfovo99} F. Dalfovo, S. Giorgini, 
L.P. Pitaevskii, and S. Stringari, Rev. Mod. Phys. {\bf 71}, 463 (1999).

\bibitem{greiner01} M. Greiner, I. Bloch. O. Mandel, 
T.W. H\"ansch, and T. Esslinger, Phys. Rev. Lett. {\bf 87}, 
160405 (2001).

\bibitem{smerzi03} A. Smerzi and A. Trombettoni, Chaos {\bf 13}, 766 (2003), 
Focus issue on ``Nonlinear localized modes: 
fundamental concepts and applications''.

\bibitem{nota} 
The ground-state wavefunction of the 
$j$th condensate
can be written as 
$\Phi_j^{var}(\vec{r})=A_j \exp{\{-m[\tilde{\omega}_x^{var} (x-x_j)^2 
+\omega_r^{var}(y^2+z^2)]/2\hbar\}}$, with $A_j$ given by the normalization 
$\int d\vec{r} (\Phi_j^{var})^2=1$.  
$\tilde{\omega}_x^{var}$ and 
$\omega_r^{var}$ are variational parameters 
determined minimizing the condensate energy 
[see, for instance, G. Baym and C.J. Pethick, 
Phys. Rev. Lett. {\bf 76}, 6 (1996)].
The chemical potential 
$\mu$ can be written as as a sum of three terms: 
$\mu=\mu_j^{kin}+\mu_j^{harm}+\mu^{int}_j$, 
where $\mu^{int}_j=g_0 N_j \int d\vec{r} (\Phi^{var}_j)^4$. The comparison 
of $\mu^{int}_j$ with the frequencies of the local potential 
$V_{local}^{(j)} = \frac{m}{2} 
[\tilde{\omega}_x^2 (x - x_j)^2 + \omega_y^2 y^2 + \omega_z^2 z^2]$
gives the effective dimensionality of the condensate, as a function of the 
number of particles $N_j$. 

\bibitem{menotti} C. Menotti, A. Smerzi and A. Trombettoni, 
New. J. Phys., in press.

\bibitem{menotti1} M. Kr\"amer, C. Menotti, L.P. Pitaevskii, 
and S. Stringari, Eur. Phys. J. D, in press (e-print: cond-mat/0305300).

\bibitem{kivshar92} Yu.S. Kivshar and M. Peyrard, Phys. Rev. A {\bf 46}, 
3198 (1992). 

\bibitem{cataliotti} F.S. Cataliotti et al., 
New J. Phys. {\bf 5}, 71 (2003); M. Kasevich et 
al., private comunication.

\bibitem{morsch02}  
O. Morsch, M. Cristiani, J.H. M\"uller, D. Ciampini, and E. Arimondo,
Phys. Rev. {\bf A66}, 021601 (2002); M. Oberthaler et al., unpublished.

\bibitem{nota-gauss} With a gaussian wavepacket of width $\sigma$,   
$\omega_{dip}^2=\frac{2 \Omega}{\hbar^2} 
\bigg( 2 K+ \frac{4 \chi N_T}
{\sqrt{\pi} \sigma} \bigg)$.

\bibitem{nota-cost} E.g., for the Thomas-Fermi function 
$f(X)=(1-X^2)^{1/\alpha}$ (with $-1 \le X \le 1$), we have:   
with $\alpha=1$, ${\cal I}_1=16/15$, 
${\cal I}_{2}=16/105$, ${\cal I}_{NL}=32/35$, 
${\cal I}_{3}=256/315$ and ${\cal I}_{4}=256/3465$; 
with $\alpha=2$, ${\cal I}_1=4/3$, 
${\cal I}_{2}=4/15$, ${\cal I}_{NL}=16/15$, 
${\cal I}_{3}=16/15$ and ${\cal I}_{4}=16/105$.  

\bibitem{nota-N} Eqs.(\ref{omega-dip},\ref{omega-quadr-TF})
can be expressed in terms of the 
atom populations $N_j$ as 
$\omega_{dip}^2= \frac{2 \Omega}{\hbar^2} 
\bigg( 2 K+ \frac{4 \chi}
{N_T} \sum_j N_j^2 \bigg)$,  
$\omega_{quadr}^2=\frac{U_\alpha \alpha (\alpha+4)}
{2\hbar^2 (\alpha+2)} \cdot 
\frac{\sum_j N_j ^{\alpha/2+1}}{\sum_j j^2 N_j^2}
\cdot \bigg( 2 K+ 4 \chi \frac{\sum_j j^2 N_j^2}
{\sum_j j^2 N_j} \bigg)$.

\bibitem{kramer02} M. Kr\"amer, L.P. Pitaevskii, and S. Stringari, 
Phys. Rev. Lett. {\bf 88}, 180404 (2002); 
C. Fort, F.S. Cataliotti, L. Fallani, F. Ferlaino, P. Maddaloni, 
and M. Inguscio, {\em ibid.} {\bf 90}, 140405 (2003).

%\bibitem{nota1} We impose
%$\int dx (\phi_G^{(j)})^2=\int dy dz (\phi_{TF}^{(j)})^2=1$.

\bibitem{trombettoni01-thesis} A. Trombettoni, Ph. D. Thesis, SISSA (2001).

\bibitem{fisher89} M.P.A. Fisher, P.B. Weichman, G. Grinstein, 
and D.S. Fisher,
Phys. Rev. B {\bf 40}, 546 (1989). 
 
\bibitem{jaksch98} D. Jaksch, 
C. Bruder, J.I. Cirac, C.W. Gardiner, and P. Zoller, 
Phys. Rev. Lett. {\bf 81}, 3108 (1998).

\bibitem{oster03} M. \"Oster, M. Johansson, and A. Eriksson, 
Phys. Rev. E {\bf 67}, 056606 (2003).

\end{thebibliography}
\end{document}